\documentclass[aps,prl,preprint,superscriptaddress,showpacs,floatfix]{revtex4-1}
\usepackage{epsfig,amssymb,amsfonts,amsmath}

\begin{document}


\title{Thermal spin injection and interface insensitivity in permalloy/aluminum metallic non-local spin valves}


\author{A. Hojem}
\author{D. Wesenberg}
\author{B. L. Zink}
\affiliation{Department of Physics and Astronomy, University of Denver, Denver, CO 80208}


\date{\today}

\begin{abstract}
We present measurements of thermal and electrical spin injection in nanoscale metallic non-local spin valve (NLSV) structures.  Informed by measurements of the Seebeck coefficient and thermal conductivity of representative films made using a micromachined Si-N thermal isolation platform, we use simple analytical and finite element thermal models to determine limits on the thermal gradient driving thermal spin injection and calculate the spin dependent Seebeck coefficient to be $-0.5\ \mu\mathrm{V}/\mathrm{K}< S_{s}<-1.6\ \mu\mathrm{V}/\mathrm{K}$.   This is comparable in terms of the fraction of the absolute Seebeck coefficient to previous results, despite dramatically smaller electrical spin injection signals.  Since the small electrical spin signals are likely caused by interfacial effects, we conclude that thermal spin injection is less sensitive to the FM/NM interface, and possibly benefits from a layer of oxidized ferromagnet, which further stimulates interest in thermal spin injection for applications in sensors and pure spin current sources. 
\end{abstract}

\maketitle


\section{Introduction}

\begin{figure*}
\includegraphics[width=4.38in]{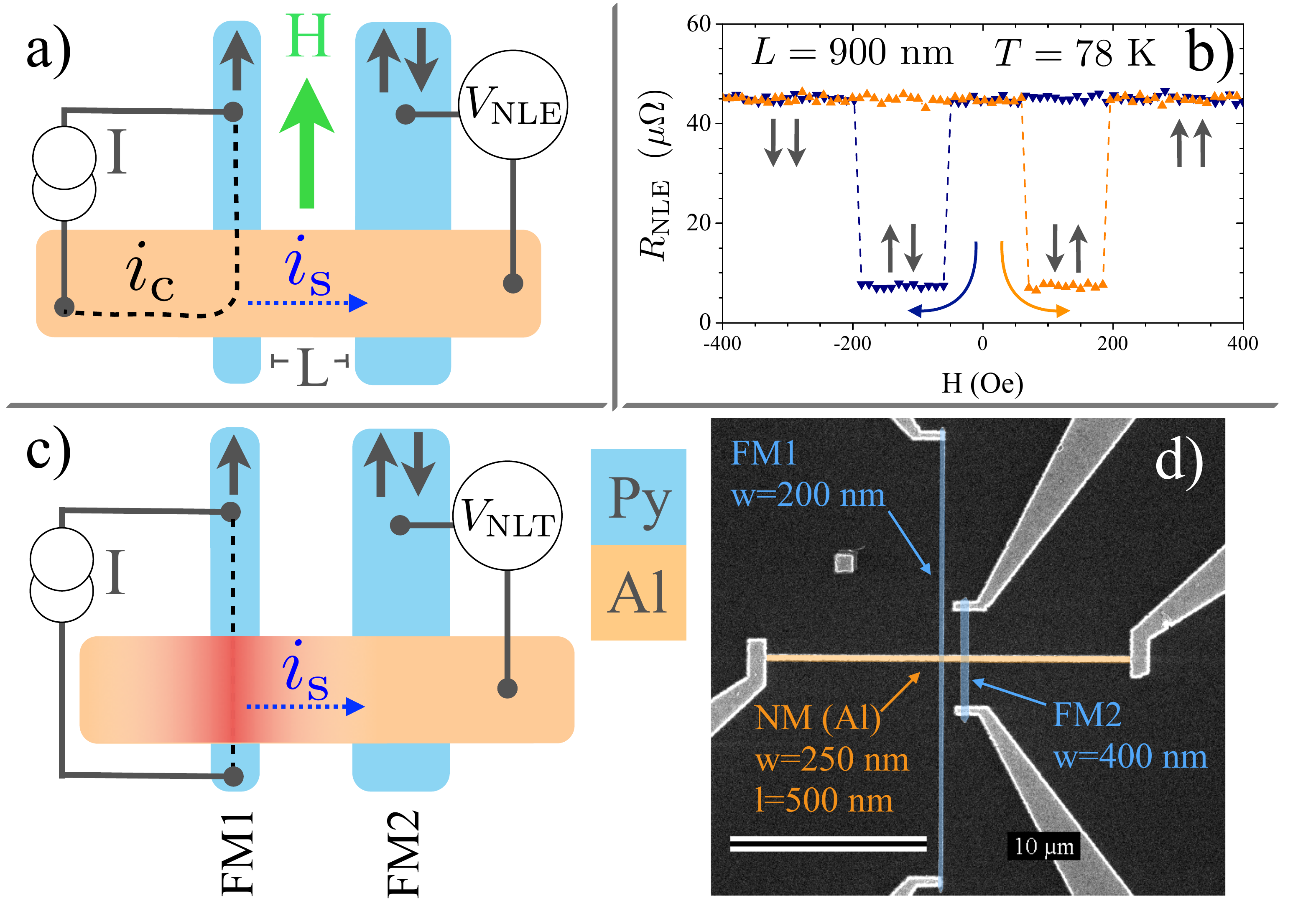}
\caption{ \textbf{a)} Schematic layout of the NLSV under electrical spin injection, where a large charge current driven through a FM nanowire creates a spin accumulation and pure spin current in a NM channel that is detected with a second FM.  \textbf{b)} The non-local resistance $R_{\mathrm{NLSV}}=V_{\mathrm{NLE}}/I$ for a $L=900\ \mathrm{nm}$ device at $78$ K, where the relative alignments of the two FM contacts are indicated with paired arrows.
 \textbf{c)} Thermal spin injection is achieved by passing current $I$ \emph{only} through FM1, creating a thermal gradient at the NM/FM interface that injects spin into the NM. \textbf{d)} False-color SEM micrograph of the nanoscale circuit defining the NLSV.  Sizes given indicate the designed widths of nanowires, measured geometries appear in Table \ref{Geometry}.}
\label{NLSVcartoon}
\end{figure*}

The non-local spin valve (NLSV), also called a lateral spin valve or spin accumulation sensor, plays an essential role in modern spintronics because of the unique ability to separate charge current from pure spin current \cite{JohnsonPRL1993,JedemaNature2001,JiAPL2004,NiimiPRL2013}.  The NLSV is formed from two ferromagnetic (FM) nanowires connected by a non-magnetic (NM) channel material with a length $L$ on the order of the spin diffusion length.  As shown schematically in Fig.\ \ref{NLSVcartoon}a), when a (charge) current is driven from the left FM contact and extracted from the nearby end of the NM channel, the spin polarization of the electrons flowing into the channel causes a transfer of angular momentum, or spin, into the NM.  This spin accumulation diffuses, decaying exponentially with distance with a spin diffusion length $\lambda_{nm}$. Note that in the ideal case no charge current is present in the NM channel where the spin accumulation leads to a pure spin current, $i_{\mathrm{s}}$.   Because of the difference in chemical potential for up and down spins, the potential difference $V_{\mathrm{NLE}}$ measured between the right FM contact and the right side of the NM channel depends on the relative alignment of the magnetization in the two FM contacts.  Dividing $V_{\mathrm{NLE}}$ and $I$ in this nonlocal geometry gives the non-local resistance resulting from electrical spin injection, $R_{\mathrm{NLE}}$, which then has the dependence on applied magnetic field, $H$, shown in Fig.\ \ref{NLSVcartoon}b).  This electrically-driven NLSV allows powerful probes of spin injection, spin accumulation, and spin transport in a wide variety of material systems \cite{TombrosNature2007,LouNatPhys2007}.  

Despite decades of study, spin transport and injection even in supposedly simple metallic systems still holds open questions and surprising results, including the role of size and material effects and nature of the injection mechanisms \cite{ObrienNatComm2014,EhrekhinskyAPL2010}.  These open questions become more urgent as industrial use of NLSV sensors for demanding magnetic field sensing applications such as read heads in magnetic recording rapidly approaches reality \cite{YamadaIEEETransMag2013}. 
Recently, thermal effects on the NLSV have proven a critical area of study, with some authors suggesting that the dominant physics driving the background resistance of the NLSV 
originates in thermoelectric effects \cite{KasaiAPL2014,HuPRB2013,BakkerPRL2010}, and others observing that significant Joule heating plays an important role in spin injection \cite{SlachterNatPhys2010,CasanovaPRB2009}.
A few groups have even shown that spin accumulation and transport in a metallic NLSV is possible by driving heat current, rather than charge current \cite{SlachterNatPhys2010,ErekhinskyAPL2012,HuNPGAM2014,HuPRB2014,YamasakiAPE2015,PfeifferAPL2015,ChoiNatPhys2015}. Such a thermal injection is shown schematically in Fig.\ \ref{NLSVcartoon}c), where current is passed only through the FM contact in order to provide a local heat source at the FM/NM interface.  If the resulting thermal gradient generates a spin accumulation in the NM and resulting spin current in the channel, the potential difference $V_{\mathrm{NLT}}$ shows a characteristic switching pattern similar to Fig.\ \ref{NLSVcartoon}b).  
This thermal generation of pure spin current, usually called the spin-dependent Seebeck effect (SDSE)\cite{BauerNatMat2012}, is still largely unexplored, and often difficult to quantify due to the need to accurately determine the thermal gradient in nanoscale structures.  There is a great deal of interest in the SDSE for applications in sensors and as a source for pure spin currents in possible spin-based logic \cite{HoffmannPRAp2015,SuSciRep2015,StampsJPhysD2014,BehinAeinNatNano2010,WolfScience01}, as well as for its role in spin-torque switching in response to fast or ultrafast laser fluence \cite{ChoiNatPhys2015}.

In this paper we present measurements of thermal and electrical spin injection and transport in all-metallic NLSVs made using permalloy (Py, the Ni-Fe alloy with $80\%$ Ni) FM and aluminum NM.  
In addition to quasi-dc measurements using the equivalent of the lock-in amplifier techniques common in the field, we fully characterize the voltage-current characteristics of the NLSV in both electrical and thermal spin injection configurations.  As discussed in detail below, this allows description of each device using a simple analytic thermal model that includes Joule heating and Peltier heating or cooling.   With knowledge of the thermal conductivity and Seebeck coefficients of representative films that we measure using our technology for thin film thermal measurements \cite{SultanJAP09,ZinkSSC10,AveryPRB2011,AveryPRB2012,AveryPRL2013,SultanPRB2013,AveryPRB2015},  we determine an upper limit on the thermal gradient driving spin injection without recourse to complicated simulations or assumptions of bulk thermal properties.   We also use a 2d finite element approach based on purely diffusive heat flow, though again informed by measured values of thermal conductivity and Seebeck coefficients, to approach a more realistic estimate of the thermal gradient and the SDSE.   The resulting SDSE coefficient for the Py/Al system at $78$ K that we report here is smaller in absolute value than previous reports using typical ferromagnets, though very comparable as a fraction of the absolute Seebeck coefficient \cite{SlachterNatPhys2010,ErekhinskyAPL2012} despite a very low efficiency of electrical injection. This suggests that thermal spin injection is far less sensitive to the nature of the FM/NM interface than its electrical counterpart and motivates broader study of the materials- and interface-dependence of thermal spin injection. 

\section{Experiment} 

 \subsection{Device Fabrication}

We fabricate NLSVs via a two-step e-beam lithography lift-off process.  Starting with silicon-nitride coated 1 cm $\times$ 1 cm Si chips with pre-patterned Au or Pt leads and bond pads, we spin an $\approx 150$ nm thick layer of PMMA that is baked for $30$ min. at $180^{\circ}$ C.  After exposure of the FM nanowire pattern using a $40$ kV SEM with the NPGS package\cite{NPGS} at a dose of $\sim600\ \mu\mathrm{C}/\mathrm{cm}^2$ and a $45\ \mathrm{s}$ development in a 1:3 MIBK:IPA solution, we deposited $100$ nm of Py from a single Ni-Fe alloy source in a load-locked UHV e-beam evaporation system at growth rates of $\sim 0.15$ nm/s.  After removal of the resist, we spin an $\approx 380$ nm PMGI spacer layer that is baked at $250^{\circ}$ for 30 min, followed by an $\approx 100$ nm thick PMMA imaging layer.  After e-beam exposure of the NM channel and lead pattern and a two-step development ($1:3:$ MIBK for $45$ s, followed by a $35$ s soak in $1:30$ solution of  2\% TMAH: IPA to form the undercut in the PMGI), we deposited a $110$ nm Al layer in a HV e-beam evaporation system at $0.2-0.5$ nm/s using a water cooled stage after a  2 minute, 50 W, -580 V RF clean process in 10 mT of Ar intended to desorb moisture from the exposed FM surface (to promote adhesion during lift-off) and potentially remove the native oxide formed on the Py nanowires.    We then remove the PMGI/PMMA resist stack via a $45$ min soak in $80^{\circ}$ C MicroChem Remover PG.  A scanning electron micrograph showing an example NLSV is shown in Fig.\ \ref{NLSVcartoon}d).

\subsection{Transport Measurements}

\begin{figure}
\includegraphics[width=3.38in]{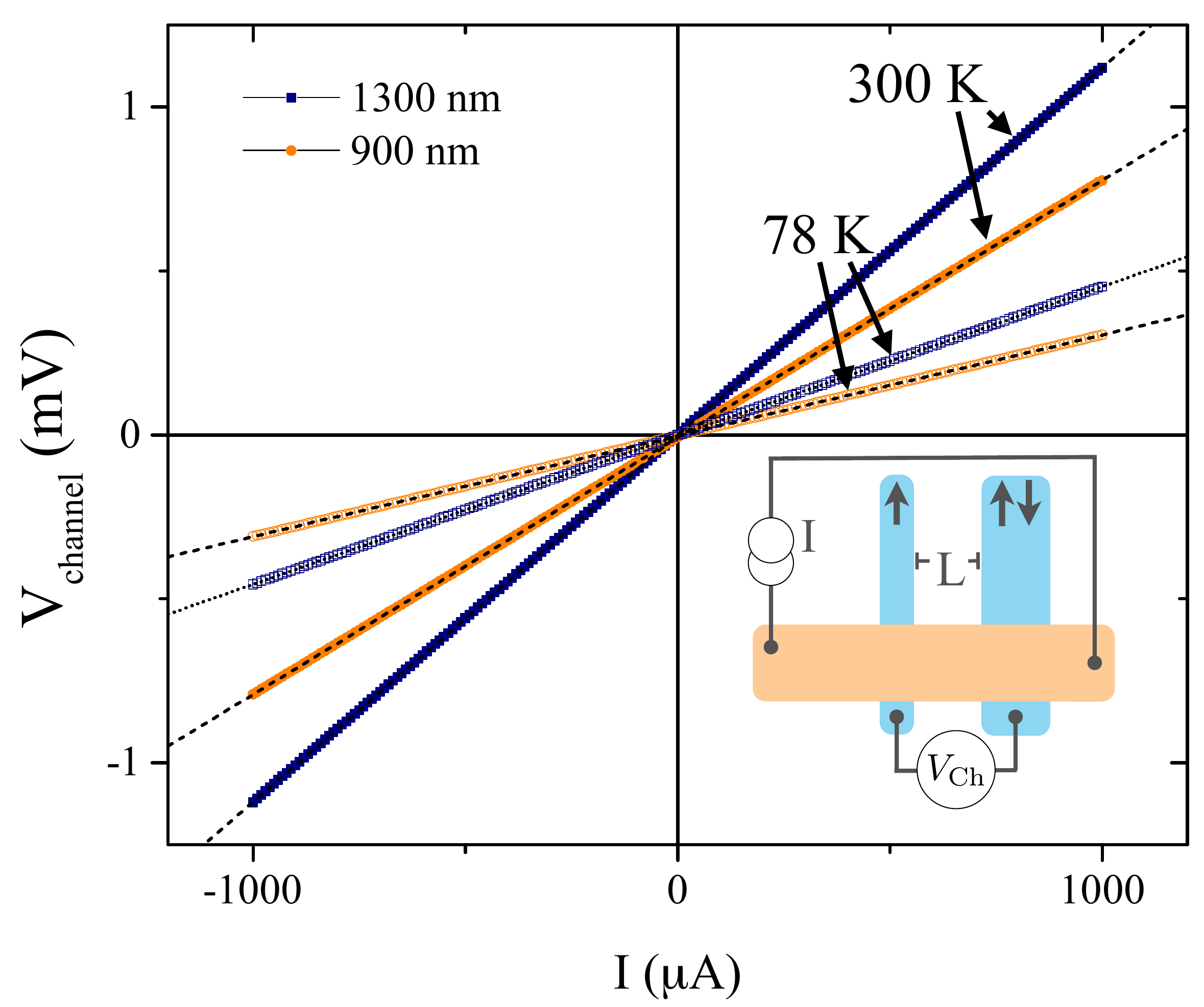}
\caption{$V_{\mathrm{channel}}$ vs. $I$ characteristics for the NM channel (contacts made as shown schematically in the inset) are highly linear across the entire range of applied $I$ in contrast to both the three-terminal contact resistance (Fig. \ref{IVs} e)) and non-local resistance measurements (Figs.\ \ref{PminAP}a) and \ref{IVs}c)).  Measurements for two NLSVs are shown for two temperatures.  Dashed lines show linear fits. }
\label{Supp1}
\end{figure}


Measurements are carried out after bolting the NLSV chip to a fully radiation-shielded gold-coated high-purity Cu sample mount installed in a sample-in-vacuum LN2 cryostat.  An open bore split-coil electromagnet allows application of fields in excess of 1000 Oe in the plane of the chip.   For the measurements described here the field is applied as shown in Fig.\ \ref{NLSVcartoon}a).  Simple resistance or non-local resistance measurements are made using the ``delta mode" function of a linked Keithley 2128a nanovoltmeter and 6220 high precision current source.  This measurement is functionally equivalent to a first-harmonic lock-in amplifier measurement \cite{DeltaMode}.  We determine IV characteristics of the NLSV in various configurations by numerically integrating differential conductance measurements made with the same system \cite{HojemThesis}.  Fig. \ref{Supp1} shows an example IV measurement of the the NM channel for the $L=900\ \mathrm{nm}$ and $L=1300\ \mathrm{nm}$ devices at both $T=78\ \mathrm{K}$ and $300\ \mathrm{K}$.  Since no FM/NM couple is in the current path in this measurement, no thermoelectric contributions are expected and indeed $V_{\mathrm{channel}}$ is highly linear for the entire range of applied $I$, as seen by the excellent agreement with linear fits shown with dashed lines.     
After all measurements are completed on a NLSV, we measure the FM and NM film thicknesses via AFM contact profilometry and the actual lateral geometry of the nanowires using SEM micrographs (see Table \ref{Geometry}).  For the devices described here, this revealed somewhat wider NM channels than intended, with widths reaching $400-450$ nm.  These measured values are used wherever geometry is needed in model calculations.

\section{Results}

\begin{figure}
\includegraphics[width=3.38in]{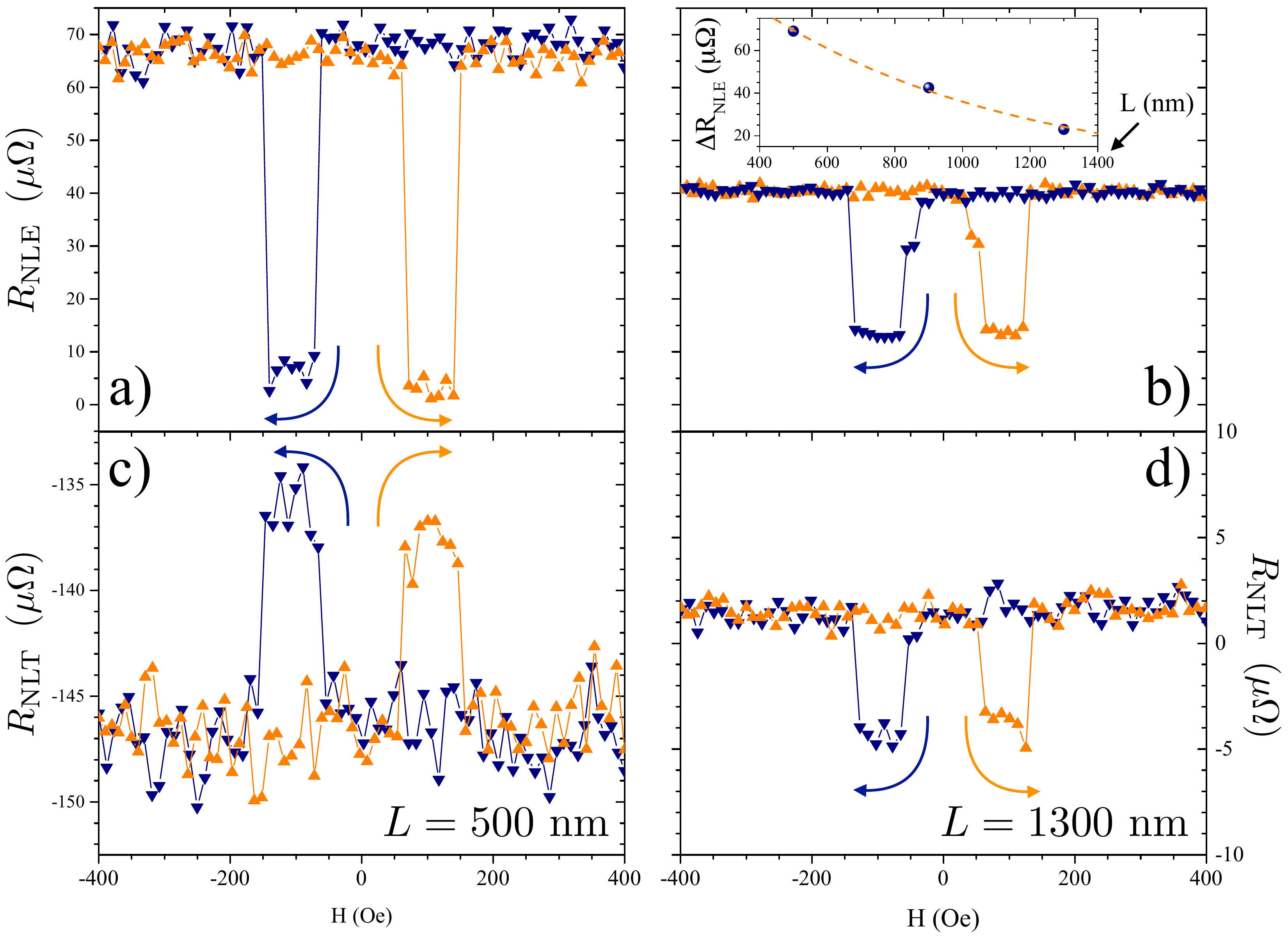}
\caption{Non local resistance signals $R_{\mathrm{NLE}}=V_{\mathrm{NLE}}/I_{b}$ in electrical (\textbf{a)} and \textbf{b)}) and $R_{\mathrm{NLT}}=V_{\mathrm{NLT}}/I_{b}$ in thermal (\textbf{c)} and \textbf{d)}) spin injection for both $500$ nm and $1300$ nm nominal FM spacing.  Inset in panel \textbf{b)} shows the electrical spin signal $\Delta R_{\mathrm{NLE}}$ vs. $L$ with the fit to the 1d spin diffusion equation.  This fit gives $\lambda_{\mathrm{nm}}=760 \pm 50\ \mathrm{nm}$.   }
\label{RvsH}
\end{figure}

Fig.\ \ref{RvsH} shows the nonlocal resistance as a function of applied field for two NLSVs with different FM spacing, $L$.  Panels \textbf{a)} and \textbf{b)} result from electrical spin injection using a bias current of $I_{b}=1\ \mathrm{mA}$ (Fig.\ \ref{NLSVcartoon}a)), while panels \textbf{c)} and \textbf{d)} current ($I_{b}=2\ \mathrm{mA}$) flows only in the FM, causing no net charge current to pass into either arm of the NM channel, but heating the FM such that a heat current forms at the FM/NM interface.  The characteristic switching clearly shows that this heating generates a spin accumulation in the NM channel that is detected after diffusing to the location of FM2.  Note however, that this quasi-dc $R$ measurement is sensitive to terms linear in $I$, where heating effects are proportional to $I^2$.  The apparent sign change in $\Delta R_{\mathrm{NLT}}=R_{\mathrm{NLT}}\left(\uparrow \uparrow \right)-R_{\mathrm{NLT}}\left(\uparrow \downarrow \right)$ is peculiar, but as is discussed in more detail below does \emph{not} indicate a sign change in the SDSE.  

As shown inset to Fig.\ \ref{RvsH}b), we use $\Delta R_{\mathrm{NLE}}=R_{\mathrm{NLE}}\left(\uparrow \uparrow \right)-R_{\mathrm{NLE}}\left(\uparrow \downarrow \right)$ to determine the spin diffusion length in the Al, $\lambda_{\mathrm{nm}}$.  Despite the RF clean step between the FM and NM depositions, comparison of the contact and spin resistances of the FM and NM (at 78 K, $R_{c}\approx 0.4\ \Omega$, $\mathcal{R}_{\mathrm{FM}}= \rho_{\mathrm{Py}}\lambda_{\mathrm{Py}}/w_{\mathrm{FM}} w_{\mathrm{nm}}\approx 16\ \mathrm{m}\Omega$, $\mathcal{R}_{\mathrm{NM}}= \rho_{\mathrm{Al}}\lambda_{\mathrm{nm}}/t_{\mathrm{NM}} w_{\mathrm{NM}}\approx 0.28\ \Omega$) indicates that $R_{c}>\mathcal{R}_{\mathrm{NM}}>\mathcal{R}_{\mathrm{FM}}$.  In this tunneling limit, the form of the 1d spin diffusion equation is \cite{TakahashiPRB2003},
\begin{equation}
\Delta R_{\mathrm{s}}=P^{2}_{I}\mathcal{R}_{\mathrm{NM}} e^{-L/\lambda_{\mathrm{NM}}}.
\label{1dSpin}
\end{equation}
The fit shown by the dashed line in the Fig.\ \ref{RvsH}b) inset gives $\lambda_{\mathrm{nm}}=760 \pm 50\ \mathrm{nm}$, which is in line with previous results for Al \cite{BassJPCM2007,ObrienNatComm2014} (for further discussion and results of fits to other 1d transport models see Appendix).  The fit, and the generally small size of $\Delta R_{\mathrm{NLE}}$, also indicates that we achieved a very low current polarization ($P_{I}=0.02$).  The value of the contact resistance area product, $R_{c}A=40\ \mathrm{m}\Omega \mu\mathrm{m}^{2}$ (from the $L=1300\ \mathrm{nm}$ NLSV), is roughly two orders of magnitude higher than seen in transparent contacts\cite{ObrienNatComm2014}, and on par with that seen in MgO tunnel barriers capable of strongly enhancing $\Delta R_{s}$\cite{FukumaNatMater2011}.  This also supports the use of the tunneling model, but with a strongly reduced $P_{I}$ due to scattering of spin introduced by a layer of oxidized Py that remains at the FM/NM interface.

\begin{figure}
\includegraphics[width=3.38in]{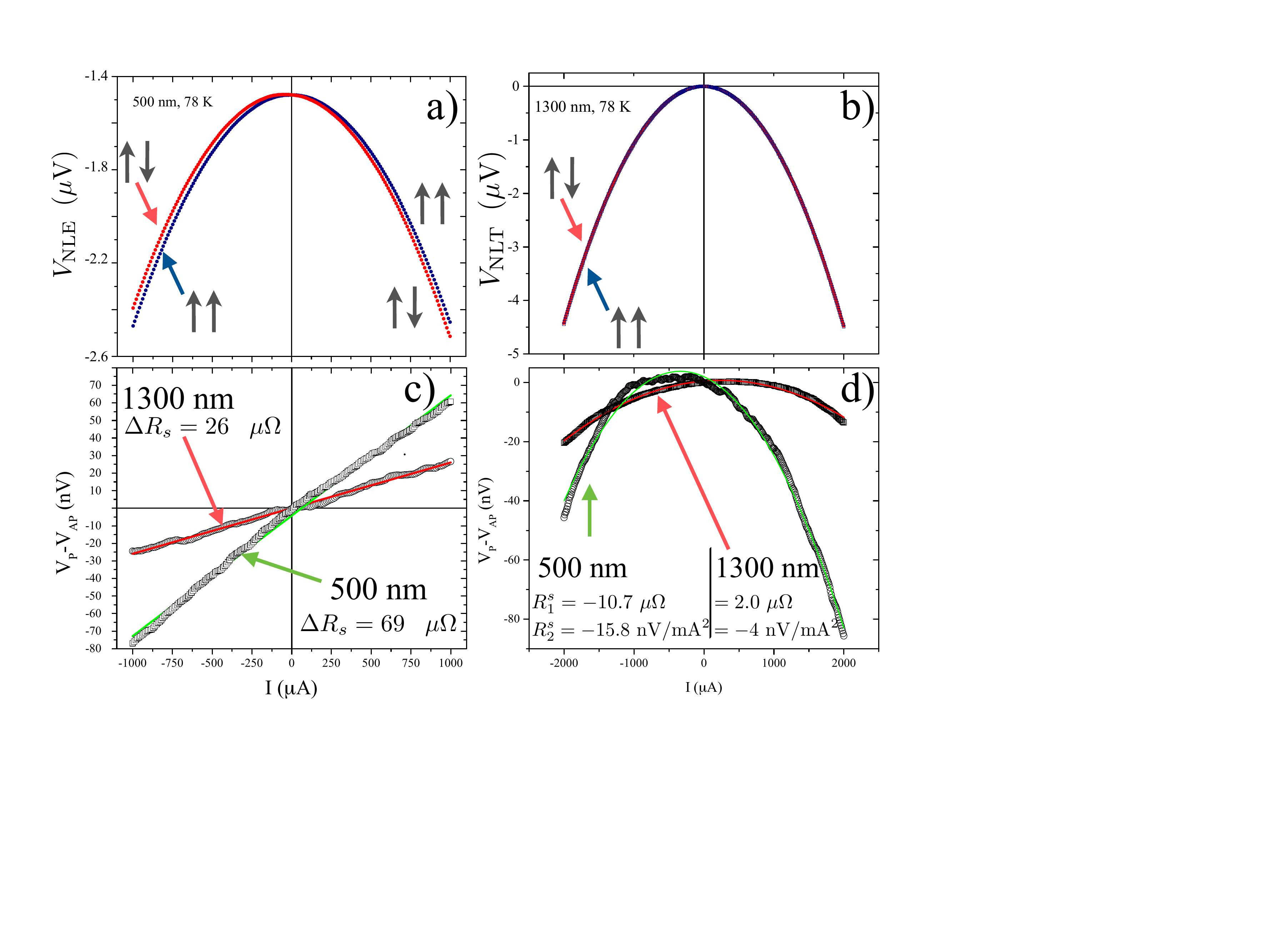}
\caption{\textbf{a)} $IV$ characteristic for the electrical spin injection configuration (Fig.\ \ref{NLSVcartoon}a) measured separately for parallel and anti-parallel states of the FM nanowires for the $L=500\ \mathrm{nm}$ device at $78$ K. \textbf{b)} The corresponding $IV$ characteristic for the thermal spin injection configuration (Fig.\ \ref{NLSVcartoon}c).  \textbf{c)} Subtraction of the parallel and antiparallel curves in \textbf{a)} gives the highly linear response of electrical spin injection, while the corresponding subtraction for thermal injection yields a spin signal dominated by the $I^{2}$ term indicating thermal generation of a spin accumulation in the NM.  In both \textbf{c)} and \textbf{d)}, data for both $L=500\ \mathrm{nm}$ and $L=1300\ \mathrm{nm}$ are shown. Fitted values of spin signal are also shown.}
\label{PminAP}
\end{figure}

Fig.\ \ref{PminAP} details the extraction of spin accumulation signals from the full IV characteristics measured in both electrical and thermal spin injection configurations.  Fig.\ \ref{PminAP}a) plots $V_{\mathrm{NLE}}$ vs. $I$ for $L=500\ \mathrm{nm}$ at $78$ K measured for two different fields, chosen based on the $R_{\mathrm{NLE}}$ vs. $H$ patterns in Fig.\ \ref{RvsH}a) to give the parallel (labeled $\uparrow\uparrow$) and antiparallel ($\uparrow \downarrow$) states of the FM nanowires.  Both curves show obvious terms $\propto I$ and $\propto I^{2}$.  The striking non-linearity is a clear indication of the importance of thermal and thermoelectric effects in this NLSV.  However,  subtracting the two curves gives the very linear response shown in Fig.\ \ref{PminAP}c) for both $L=500\ \mathrm{nm}$ and $L=1300\ \mathrm{nm}$, where the slope matches the spin signal seen in $R_{\mathrm{NLE}}$ vs. $H$.  Fig.\ \ref{PminAP}b) and d) show similar plots for thermal spin injection ($V_{\mathrm{NLT}}$) measured at the same temperature over a wider $I$ range.  As expected $V_{\mathrm{NLT}}$ is predominantly $\propto I^{2}$, and the difference between parallel and antiparallel configurations (Fig. \ref{PminAP}d) retains a large $\propto I^{2}$ component.  Lines in Fig. \ref{PminAP}d) are fits to $V_{\mathrm{P}-\mathrm{AP}}=R^{s}_{1} I + R^{s}_{2} I^{2}$.  As discussed further below, the $R^{s}_{2}$ provides the same information as the second-harmonic lock-in signal in previous work \cite{SlachterNatPhys2010}, and is the evidence of thermally-generated spin accumulation in the NLSV.  The physics of the $R^{s}_{1}$ term is less clear, though this term was also seen in the original report of the SDSE \cite{SlachterNatPhys2010}.  In fact, the size of  $R^{s}_{1}$ and $R^{s}_{2}$ shown in Fig. \ref{PminAP}d) for $L=500\ \mathrm{nm}$ is nearly the same as the results in \cite{SlachterNatPhys2010}.  However, this does not necessarily imply a similar SDSE coefficient, since the thermal profile in the NLSV must be determined and will certainly depend on the detailed geometry and materials in each device.  We also point out that the difference in sign in $R^{s}_{1}$ between the $500\ \mathrm{nm}$ and $1300\ \mathrm{nm}$ devices entirely explains the sign change of $\Delta R_{\mathrm{NLT}}$ apparent in Figs.\ \ref{RvsH}c) and d) and clarifies that this is not related to the SDSE.  Recent electrical injection experiments in the wiring configuration of Fig.\ \ref{NLSVcartoon}b for a Py/Cu NLSV with Al$_{2}$O$_{3}$ tunnel barriers showed a spin accumulation signal that was interpreted as evidence of a non-uniform spin injection across the contact.\cite{ChenJAP2015}  A similar mechanism could well explain our $R^{s}_{1}$, but requires further study to conclusively discuss.

\section{Discussion}


Accurately determining the thermal gradient generated in any nanoscale metallic device is a serious challenge.  Even if complicated 3d finite element analysis (FEM) is used, having accurate values of thermal properties for the thin film constituents of the devices is important, and the role of interfaces for electron, phonon, and spin transport is difficult to quantify without great effort \cite{CahillAPR2014,CahillJAP2003}.  Furthermore, typical codes describe only diffusive heat transport, ignoring ballistic or quasi-ballistic phonon transport that is known to play a role in nanoscale metallic features on insulating substrates \cite{SiemensNatMat2010}.  In fact the previously common view that only phonons of quite short wavelength and mean-free-path dominate heat transport in bulk materials at room temperature is now understood to be incorrect, with more and more quantitative measurements showing large contributions to heat flow from parts of the phonon spectrum ignored in typical FEM simulations \cite{RegnerNMTE2015,LarkinPRB2014,RegnerNatComm2013,SultanPRB2013,JohnsonPRL2013,MinnichPRL2011}.    These issues suggest that truly quantitative determination of the SDSE coefficient will be challenging and some level of disagreement between experimental groups should be expected, a situation familiar to the spintronics community.  

We therefore clarify that the main result of this study requires no complicated or controversial calculations of thermal gradients.  First consider that the spin signal due to electrical spin injection in the NLSV first used for the SDSE measurement by Slachter, et al. was  $\Delta R_{s}\approx10\ \mathrm{m}\Omega$ where the thermal injection signal as discussed earlier was $R^{s}_{2}= -16\ \mathrm{nV}/\mathrm{mA}^2$.  In the NLSV devices described here we achieved the same thermal spin signal $R^{s}_{2}$ despite an electrical spin signal of only $\Delta R_{s}\approx 70\ \mu\Omega$, a factor of more than $100$ smaller.  We can also use a simple 1d Valet-Fert model for spin diffusion to make a more fair comparison of spin accumulation at the injection site between devices and injection techniques.  This suggests that Slachter et al.'s 100 nm asymmetric NLSV where ion milling was used to remove Py oxide at the interfaces showed thermal spin accumulation of $<0.2$ \% of electrical spin accumulation at the same applied current.  Our NLSVs, where Py oxide likely remains at the interface, show  similar thermal spin accumulation but dramatically smaller electrical spin accumulation so that the ratio is $> 0.15$ \%.  As discussed further below, this suggests that thermal spin injection is much more tolerant of imperfect interface quality, and in fact may be enhanced by the presence of an oxidized Py layer.     

\begin{figure}
\includegraphics[width=3.2in]{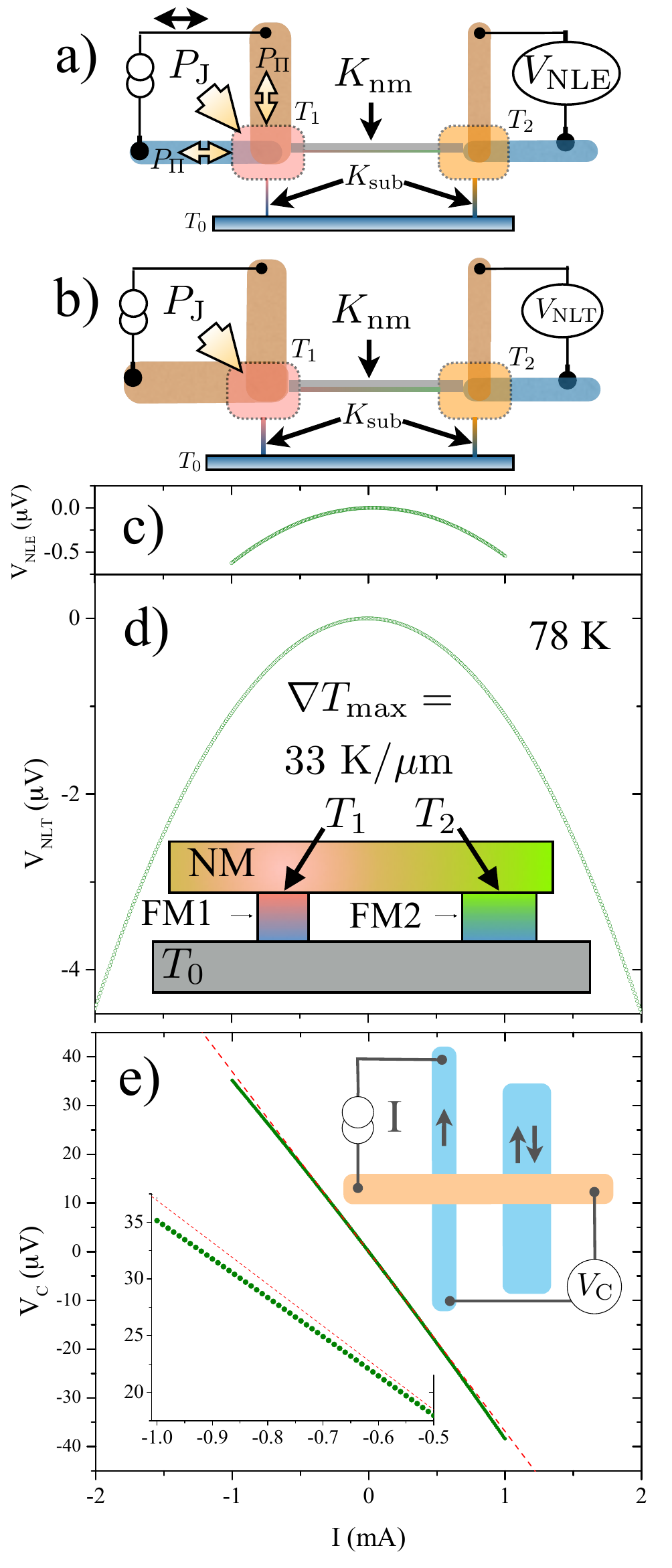}
\caption{\textbf{a-b)} Two-body thermal models used to analytically model the $T$ profile in the devices. \textbf{c-d)} Resulting $IV$ curves show significant curvature as a result of heating and thermoelectric effects.  Data is shown for the $L=1300\ \mathrm{nm}$ NLSV at $78$ K, but similar curvature is seen at room $T$ and for other devices.  \emph{Inset:} The simplified thermal profile used to estimate a maximum possible $\nabla T$ of $33$ K/micron from our data.  \textbf{e)} The three-terminal contact resistance (shown schematically in upper inset) $IV$ characteristic shows small but clearly measurable non-linearity (lower inset). }
\label{IVs}
\end{figure}

We now consider two techniques for estimating the thermal gradient driving the SDSE in our NLSVs. 
The first is a simple analytic technique using the two-body thermal models shown in Fig. \ref{IVs}a) and b). 
Here we assume the two FM/NM junctions equilibrate to two different temperatures in steady state, $T_{1}$ and $T_{2}$, that both junctions are connected to thermal ground (the substrate held at $T_{0}$) via the same thermal conductance $K_{\mathrm{sub}}$, and that heat can flow between the two junctions via thermal conductance $K_{\mathrm{nm}}$.  This model is shown schematically for electrical spin injection in Fig. \ref{IVs}a).  Note that truly ascribing physical meaning to the parameters in this simple model is difficult.  For example one would normally expect that the NM channel in a typical NLSV would be coupled to the bath (substrate) with approximately the same thermal conductance as the junctions, though all these features are on the size scale where decoupling from the phonons responsible for heat-sinking the metal structures can lead to larger heating effects and counterintuitive behavior \cite{SiemensNatMat2010}.  

As already noted by other groups  \cite{BakkerPRL2010,KasaiAPL2014,CasanovaPRB2009}, when current is driven into the injector FM and out of one arm of the NM channel,  Joule heating in this current path is accompanied by either cooling or heating due to the Peltier effect.  Whereas Joule heating, $P_{\mathrm{J,}i}=I^2R_{\mathrm{eff}}$,  is always positive, the Peltier term, $P_{\Pi_{\mathrm{rel}}}=I\Pi_{\mathrm{rel}}$, is either positive or negative.   The sign of the Peltier term depends on the direction of applied current, the geometric arrangement of the two metals with respect to this current flow, and the difference in the absolute Peltier coefficients of the two materials (written here simply as the relative coefficient $\Pi_{\mathrm{rel}}$). Furthermore, via Onsager reciprocity\cite{DejenePRB2014,AveryPRL2013}, $\Pi_{\mathrm{rel}}=\alpha_{\mathrm{rel}}T_{0}$, where we use the substrate temperature since deviation in $T$ even by several Kelvin makes a negligible change in the Peltier power at the $T$ studied here.

\begin{figure}
\includegraphics[width=3.38in]{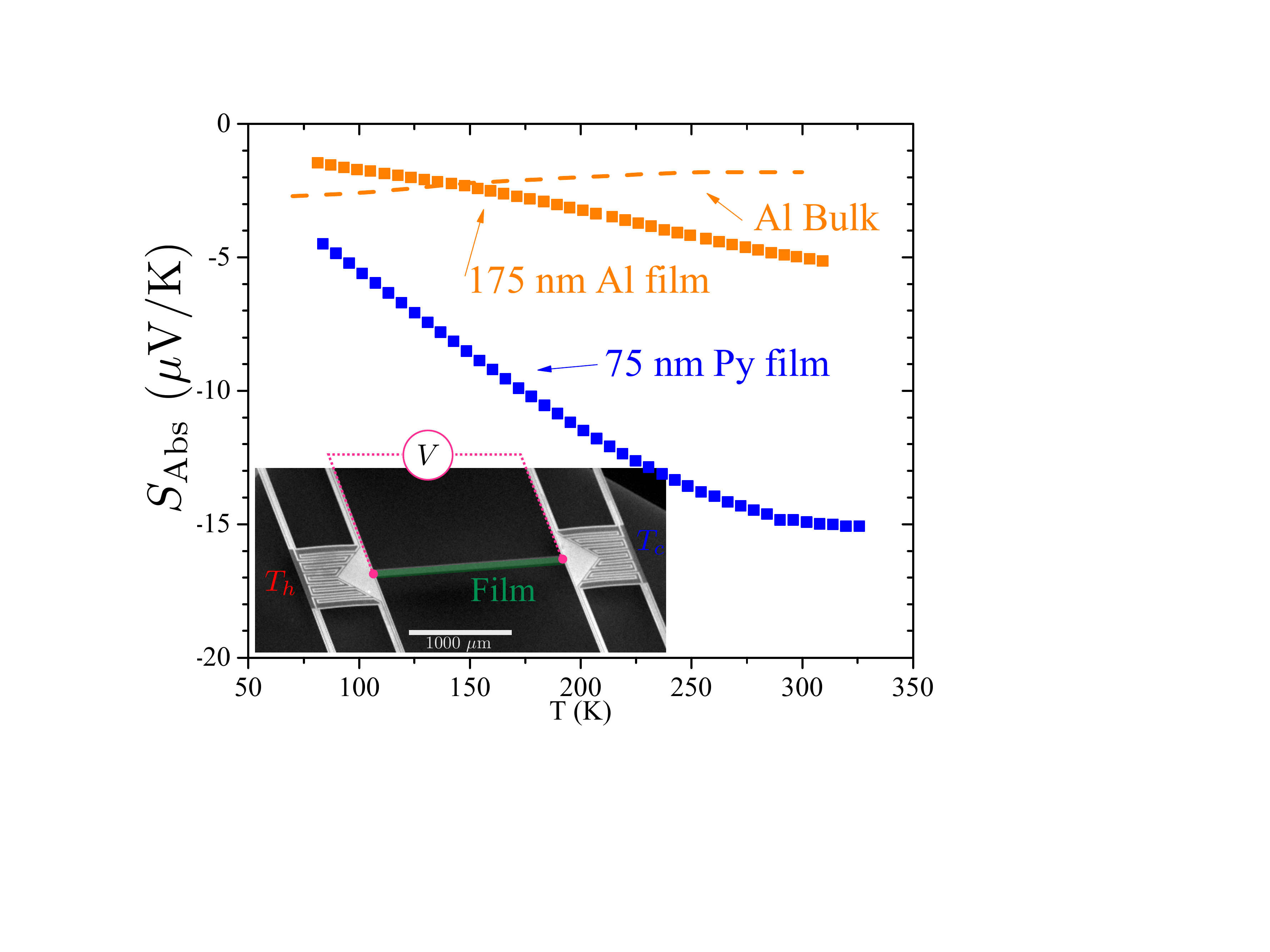}
\caption{ Measured Seebeck coefficients for the constituent thin films vs. $T$.   Each film was deposited on a thermal isolation platform, and the measured Seebeck coefficient is relative to Cr/Pt leads.  The estimated lead contribution has been subtracted here, so that this plot compares estimated \emph{absolute} Seebeck coefficients.   \emph{Inset:} SEM micrograph of the thermal isolation platform we use for thermal properties measurements.}
\label{Seebeck}
\end{figure}

The schematics in Fig.\ \ref{IVs}a) and b) lead to a coupled system of equations that can be compared to fits of the full $IV$ characteristics in the configurations shown in Figs.\ \ref{NLSVcartoon}a) and b) and Fig.\ \ref{IVs}e).  Each of these measurements contains terms proportional to $I$ and to $I^{2}$ and are fit to:
\begin{eqnarray}
V_{\mathrm{NLE}} & = & A_{1}I + A_{2}I^{2} \label{FitEqVnle}\\
V_{\mathrm{NLT}} & = & B_{1}I + B_{2}I^{2} \label{FitEqVnlt}\\
V_{\mathrm{C},1} & = & C_{1}I + C_{2}I^{2} 
\label{FitEqVc}
\end{eqnarray}
Collecting terms in the corresponding thermal model that are proportional to $I$ and $I^{2}$ and solving these systems of equations yields expressions for the thermal parameters (as shown in Appendix A).  With certain assumptions listed below we can then calculate the temperature difference between the heated region of junction 1 and the substrate in thermal spin injection, $\Delta T^{t}_{1}$.  This is the critical value needed to calculate the SDSE coefficient, $S_{s}$. First we assume that the parameter $K_{\mathrm{nm}}$ is given by the thermal conductance of the normal metal nanowire itself (ignoring any heat transported by the underlying substrate) and use the Wiedemann-Franz law to determine this $K_{\mathrm{nm}}$ from the measured resistance of the channel, $R_{\mathrm{nm}}$, 
\begin{equation}
K_{\mathrm{nm}}=\frac{L_{\mathrm{Al}} T_{0}}{R_{\mathrm{nm}}}.
\label{Knm}
\end{equation}  
Here we take the value of the Lorenz number, $L_{\mathrm{Al}}=2.0\times10^{-8}\ \mathrm{W}\Omega/\mathrm{K^{2}}$  from a measurement of a similar Al thin film made using our micromachined thermal isolation platform \cite{AveryPRB2015}. 
Next we assume that both the injection and detection FM/NM arms of the NLSV have the same value of $\alpha_{\mathrm{rel}}$.  Though thermopower is often assumed to be independent of geometry, this is only strictly true in the case where thermal gradient is simply aligned with the sample and in the regime where size effects cannot play a role.  Nanoscale metal features are not always in this simple limit \cite{StrunkPRL98,SzakmanyIEEETransNano2014}, so our model could be improved using actual measurements of Seebeck effects in nanowires of the same dimension as used in the NLSV.  Since these measurements are not possible for the current devices, we instead take a value of the relative Seebeck coefficient at $78$ K again from measurements of representative films made using thermal isolation platforms.  

Seebeck coefficient data is shown in Fig.\ \ref{IVs}f), where we present estimated absolute Seebeck coefficient as a function of $T$ for both Al and Py films.  These measurements are made on thin films deposited on a patterned $500$ nm thick suspended silicon-nitride membrane with integrated heaters, thermometers, and electrical contacts.  Application of a temperature difference $\Delta T=T_{H}-T_{c}$ generates a voltage across the film due to the Seebeck effect, $V$, giving the relative Seebeck coefficient, $\alpha_{\mathrm{rel}}=V/\Delta T=\alpha_{\mathrm{abs}}-\alpha_{\mathrm{lead}}$.  Note that both measurements are made with the same lead material, so the determination of $\alpha_{\mathrm{abs}}$ (which adds some uncertainty) is not necessary to determine the value needed for NLSV modeling, $\alpha_{\mathrm{rel}} =\alpha_{\mathrm{Al}}-\alpha_{Py}$.  More details about Seebeck measurements made with our thermal isolation platforms are available elsewhere \cite{AveryPRB2011,AveryPRB2012,AveryPRL2013,MasonThesis}.

With these assumptions we can write, 
\begin{equation}
K_{\mathrm{sub}}= \left( \frac{C_{2}}{A_{1}}\frac{\alpha_{\mathrm{rel}}T_{0}}{R^{e}_{\mathrm{eff}}} -1\right) K_{\mathrm{nm}},
\label{Ksub} 
\end{equation}
where here we  use $R^{e}_{\mathrm{eff}}=\alpha_{\mathrm{rel}}T_{0} (A_{2}/A_{1})$ for the contact resistance measurement to determine $K_{\mathrm{sub}}$.  The temperature rise at the injector junction is then
\begin{equation}
\Delta T^{t}_{1}=\left( \frac{B_{2}}{\alpha_{\mathrm{rel}}} \frac{\left(K_{\mathrm{sub}}+2K_{\mathrm{nm}}\right)}{K_{\mathrm{nm}}}\right ) \left[ 1-\frac{A_{1}K_{\mathrm{sub}}}{\alpha_{\mathrm{rel}}^{2}T_{0} }\right] I^{2}.
\label{DeltaT1}
\end{equation}
The $B_{2}$ term enters from use of $R^{t}_{\mathrm{eff}}=\alpha_{\mathrm{rel}}T_{0} (B_{2}/A_{1})$ to account for the different effective resistance when current flows only through FM1.  

\begin{table}
\begin{tabular}{| | c | c | c ||}
\hline\hline
 & $500\ \mathrm{nm}$  & $1300\ \mathrm{nm}$ \\
 \hline
 $A_{1}$  & $3.9\ \mu\Omega$ & $39.34\   \mu\Omega$\\
 $A_{2}$    & $-0.984\ \mathrm{V/A}^{2}$& $-0.586\ \mathrm{V/A}^{2}$ \\
 \hline
 $B_{1}$ & $-146.95\ \mu\Omega$ \ & $-11.43\ \mu\Omega$\\
$B_{2}$  & $-1.498\ \mathrm{V/A}^{2}$  &  $-1.112\ \mathrm{V/A}^{2}$  \\
\hline
 $C_{1}$  & -- & $-36.76\ \mathrm{m}\Omega$ \\
$C_{2}$    &  $-1.76^{\dagger}\ \mathrm{V/A}^{2}$ & $-1.66\ \mathrm{V/A}^{2}$ \\
\hline
\hline
$\Delta T^{t}_{1}\ (2\ \mathrm{mA})$  & $5.3$ K  & $3.3$ K \\
$\nabla T^{t}_{1}\ (2\ \mathrm{mA})$ & 53 K/$\mu$m & 33 K/$\mu$m \\
\hline
$S_{s}$ & $-0.46\ \mu\mathrm{V}/\mathrm{K}$ & $-0.53\ \mu\mathrm{V}/\mathrm{K}$ \\
\hline\hline
$\Delta T^{\mathrm{FEM}}_{1}\ (2\ \mathrm{mA})$  & $3.9$ K  & $5.4$ K \\
$\nabla T^{\mathrm{FEM}}_{1}\ (2\ \mathrm{mA})$ &  15 K/$\mu$m & 23 K/$\mu$m \\
\hline
$S_{s,\mathrm{FEM}}$ & $-1.6\ \mu\mathrm{V}/\mathrm{K}$ & $-0.77\ \mu\mathrm{V}/\mathrm{K}$ \\
\hline\hline
\end{tabular}
\caption{Fitting parameters as defined in Eqs.\ \ref{FitEqVnle}-\ref{FitEqVc} and resulting temperature difference, and absolute values of thermal gradient from the analytic thermal model, ( $\Delta T^{t}_{1}$ and $\nabla T^{t}_{1}$) and resulting lower limit on SDSE coefficient, $S_{s}$ compared to temperature difference, thermal gradient, and SDSE coefficient from FEM modeling, ( $\Delta T^{\mathrm{FEM}}_{1}$, $\nabla T^{\mathrm{FEM}}_{1}$, and $S_{s,\mathrm{FEM}}$)  .    $\dagger$: Value calculated from model assuming the same value of $K_{\mathrm{sub}}$ for both devices.}
\label{table}
\end{table}

$\Delta T^{t}_{1}$ for the two NLSVs for two different currents are shown in Table\ \ref{table}, and indicate the NLSV junctions heat by several Kelvin during operation in thermal injection.  The SDSE coefficient, $S_{s}$, following \cite{SlachterNatPhys2010} is
\begin{equation}
S_{s}= \frac{V_{s}}{\nabla T \lambda_{\mathrm{FM}} R_{\mathrm{mis}}}, 
\label{sdse}
\end{equation}
where $V_{s}=-\mu_{s}/e$ is the spin accumulation at the injection junction (FM1), and $R_{\mathrm{mis}}=\mathcal{R}_{\mathrm{NM}}/(\mathcal{R}_{\mathrm{NM}}+\left(\mathcal{R}_{\mathrm{FM}}/1-P_{I}^2\right)$ is always $\cong 1$ for these metallic NLSVs.  
To estimate the SDSE Coefficient, $S_{s}$, we need to determine a thermal gradient at the injection site from our temperature difference.   For the analytic model we assume the highly simplified situation shown schematically inset in Fig. \ref{IVs}d), where the temperature $T^{t}_{1}=T_{0}+\Delta T^{t}_{1}$ is the effective temperature of the interface between FM and NM, and apply the 1d heat flow equation across the FM with the boundary conditions of $T_{0}$ and $T^{t}_{1}$, which gives a linear thermal gradient in the FM.  The resulting $\nabla T^{t}_{1}$ for two applied currents is also shown in Table \ref{table}, and is comparable to that calculated in other work for large $I$ \cite{SlachterNatPhys2010,ErekhinskyAPL2012,HuNPGAM2014}.  Note that this simple assumption amounts to the limit where the NM channel can only exchange heat with the top surface of each FM contact, and is most likely not physically accurate.  However, it does provide an estimate for the \emph{largest} absolute value of gradients possible in our structure because it ignores heat-sinking by the NM channel which will lower $\nabla T$ at the interface.

\begin{figure}
\includegraphics[width=3.38in]{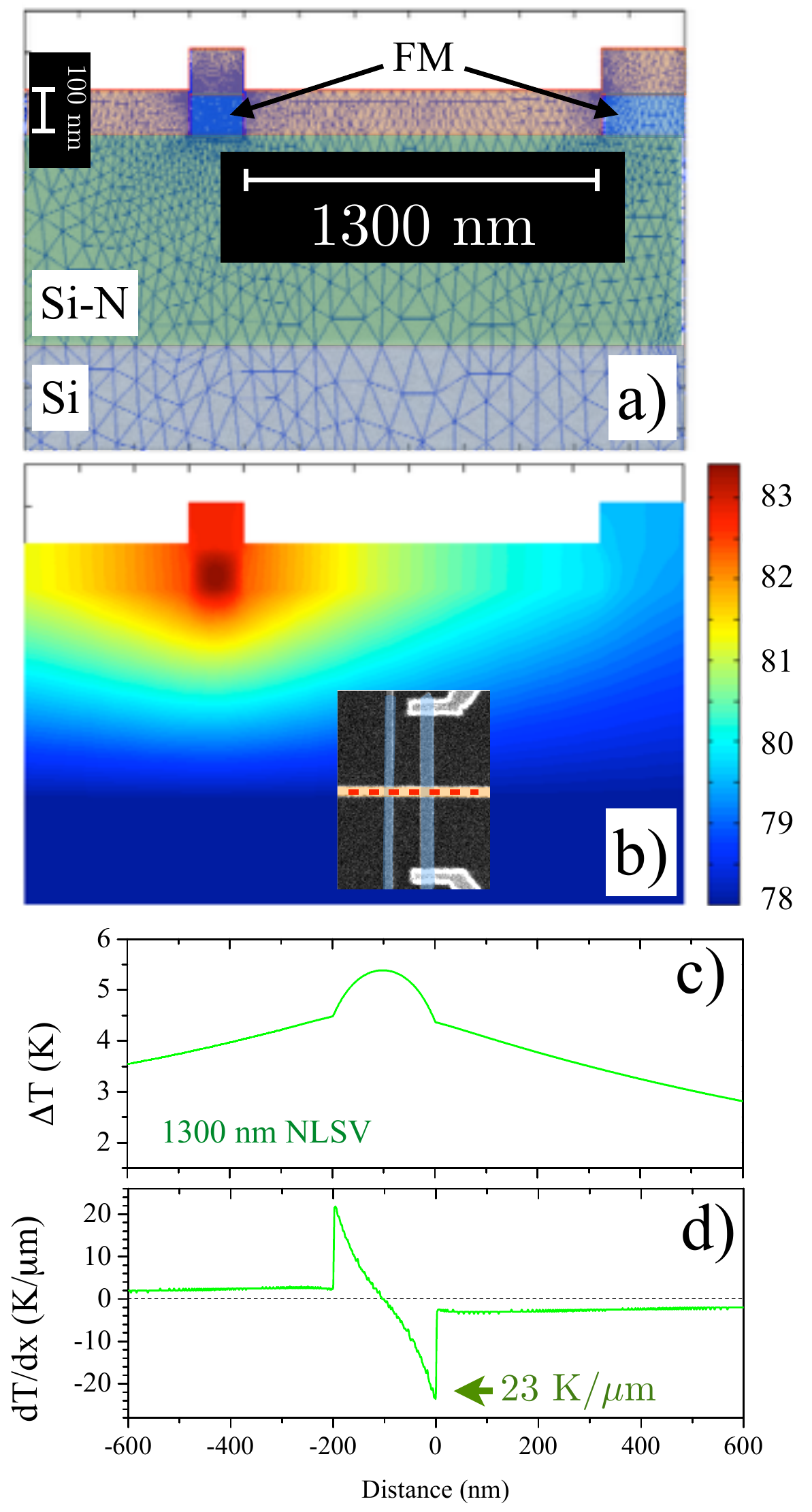}
\caption{\textbf{a)} 2d geometry and mesh used for FEM thermal calculations.   \textbf{b)} Thermal profile resulting from heat dissipated in FM1 chosen to give the correct $\Delta T$ at FM2. \emph{Inset:} Dashed red line shows the region of the 2d cross-sectional slice used for the FEM model.   \textbf{c-d)} Resulting $T$ and $dT/dx$ profiles for the $L=1300\ \mathrm{nm}$ NLSV at the height $\approx50\ \mathrm{nm}$ above the substrate at the peak of the broad maximum in $dT/dx$.}
\label{FEM}
\end{figure}

The opposite limit is described by a purely diffusive heat flow model that allows exchange of energy between elements in the real geometry of the device.    3d finite-element modeling (FEM) calculations that couple the heat, charge, and spin degrees of freedom to calculate $\nabla T$ in this limit have already been demonstrated \cite{FlipseNatNano2012,SlachterNatPhys2010,BakkerPRL2010}.  
The second thermal modeling approach we take is a simple FEM calculation focusing only on the thermal degrees of freedom, and taking 2d ``slices" through the device structure in critical areas.  Similar 2d FEM codes have been frequently used to describe heat flow in micro- and nanomachined calorimeters \cite{LeeTCA2015,BaldasseroniRSI2011,AnahoryTCA2010,simulation}.   
We performed 2d FEM using a common commercially available software package \cite{MatLab}.  This allows solution of the 2d heat flow equation (for our purposes limited to steady-state):
\begin{equation}
\frac{\partial}{\partial x} \left( k_{\mathrm{2D}}\left(x,y\right)\frac{\partial T\left(x,y\right)}{\partial x}\right)+\frac{\partial}{\partial y} \left( k_{\mathrm{2D}}\left(x,y\right)\frac{\partial T\left(x,y\right)}{\partial y}\right)=P_{\mathrm{2D}}\left(x,y\right),
\end{equation}
where $k_{\mathrm{2D}}=k \cdot t$ with $k$ the thermal conductivity (in $\mathrm{W}/\mathrm{m} \mathrm{K}$) of the constituent materials shown in Fig.\ \ref{FEM}a) and $t$ is a uniform thickness (here $450$ nm) of the hypothetical cross-section.  As long as the heat flow is dominated by the bulk substrate so that in-plane thermal transport is negligible on long length scales, such a model gives a reasonable estimate of the thermal gradient at the FM/NM interface.  To match our experimental conditions (sample in vacuum, with substrate clamped at the bottom to a thermal bath), we choose the Dirichlet boundary condition at the base of the Si substrate (fixing $T=78$ K), and Neumann boundary conditions elsewhere with no radiative or convective heat flow.   

Values of the thermal conductivity of the metallic nanowires are determined in the same fashion as for the analytic model (the WF law with modified $L$ for Al and using measured values for similar thickness of Py).  For the Si-N underlayer, which is critical for realistic modeling, we take the value $\sim 3\ \mathrm{W}/\mathrm{m} \mathrm{K}$ that we measure frequently for this Si-N using the suspended Si-N platforms \cite{SultanPRB2013}, and use literature values for Si thermal conductivity ($2000\ \mathrm{W}/\mathrm{m} \mathrm{K}$  at 78 K) \cite{GlassbrennerPR1964}.  For simplicity we use temperature-independent thermal conductivity (since most of these materials have $k$ that varies slowly if at all over the few-kelvin range of heating we expect), and also make the simplifying assumption that all Joule heat is dissipated in the FM1 nanowire.  In the case of Py (a high electrical resistivity alloy) and Al (a potentially low conductivity metal) with a truly clean interface and bulk-like values of $\rho$ this would likely be a poor assumption.  However, the reduced size, impurity and roughness, and likelihood of less-than ideal contact all suggest that modeling this limit could be more realistic.  Any spreading of the applied current to FM1 into the NM channel would cause some amount of the dissipated power to occur also in the NM, which would serve to reduce the thermal gradient calculated at the FM/NM interface.  This would then increase the value of $S_{s}$ estimated from the FEM model.  Overall, this challenge falls in the realm of the difficulty all groups have with taking interface heat flow and thermal properties correctly into account when performing thermal modeling.  

We set $P_{\mathrm{2d}}$ dissipated in FM1 by matching the temperature difference to that required to generate the measured voltage response at the FM2/NM thermocouple.  The FEM problem is then solved using an adaptive mesh with $> 5000$ nodes (as shown in Fig.\ \ref{FEM}a)). The resulting solution for $T \left(x,y\right)$ is shown in Fig.\ \ref{FEM}b), and this solution is plotted for the height midpoint of the NM channel as a function of length along the channel in Fig.\ \ref{FEM}c).  The numerical derivative of this curve gives the thermal gradient $dT/dx$ as a function of $x$ as shown in Fig.\ \ref{FEM}d).  
As expected this indicates somewhat smaller thermal gradients in the FM within one spin diffusion length of the interface compared to the analytic model.   Note also that the thermal gradient vector at the FM/NM channel interface points toward the FM (in the negative $x$ direction) for this device. The same operating conditions discussed above for the $L=1300\ \mathrm{nm}$ device at $78$ K give $\nabla T_{\mathrm{FEM}}=23\ \mathrm{K}/\mu \mathrm{m}$.  The same procedure applied to the $500$ nm geometry gives a yet lower thermal gradient, which most likely indicates breakdown in the assumptions, and possibly that the relative Seebeck coefficients or thermal conductivities are in fact not the same between these devices.

To calculate $S_{s}$ we then assume a value of $\lambda_{\mathrm{FM}}=5\ \mathrm{nm}$ for Py for easiest comparison to other work, though note that variation in this value directly affects $S_{s}$ and that our results would be best discussed as the product $S_{s}\lambda_{\mathrm{FM}}$.  Finally, we determine $V_{s}$ via solution of the Valet-Fert equation using measured $V_{s}=R^{s}_{2}I^{2}$ at the detector junction,  $\lambda_{\mathrm{nm}}$, and $L$ for each NLSV.  The result (for $L=1300$ nm) is $S_{s}=-0.5\ \mu\mathrm{V}/\mathrm{K}$ (from the analytic method) and $S_{s}=-0.77\ \mu\mathrm{V}/\mathrm{K}$ (from the FEM method) for our Py/Al at $78$ K.  This absolute value is somewhat smaller than other reports, which range from $S_{s}=-3.8\  \mu\mathrm{V}/\mathrm{K}$ for Py/Cu at $300$ K in the original report \cite{SlachterNatPhys2010}, to as large as $S_{s}=-72\ \mu\mathrm{V}/K$ for CoFeAl/Cu also at $300$ K where the strong enhancement is believed to relate to formation of a half-metallic phase in the CoFeAl film \cite{HuNPGAM2014}.  However, viewed as a fraction of the T-dependent total absolute Seebeck coefficient of Py, $S^{Py}_{\mathrm{abs}}$, in order to compare across the different measurement temperatures, our value $S_{s}/S^{Py}_{\mathrm{abs}}=0.12-0.3$ is closer to (and perhaps even in excess of) that seen in other Py devices $S_{s}/S^{Py}_{\mathrm{abs}}=0.19$ \cite{SlachterNatPhys2010}.


It is quite remarkable that the size of the thermal spin injection signals corresponds to this very significant degree of polarization of the Seebeck coefficient when the current polarization, $P_{I}=0.02$, determined from the size and $L$ dependence of the electrical spin signal is so low.  As stated above, we attribute the low electrical injection signals and $P_{I}$ to loss of spin information as the spin polarized electrons are injected in the NM.  In theory the low $P_{I}$ could indicate a reduced spin polarization in the bulk of the Py itself, though films made from this source in this chamber have historically not shown dramatically reduced values of $M_{s}$, AMR, or of course Seebeck coefficient \cite{ZinkPRB2015,WesenbergPRBinprep,AveryPRB2011}.  The most likely cause for the reduced electrical spin injection is the formation of oxidized permalloy at the FM/NM junction that was not fully removed by the RF cleaning step before Al deposition.  Native permalloy oxides can be complicated chemically and magnetically \cite{FitzsimmonsPRB2006}, though typically are not seen to develop long-range magnetic order above $\sim30$ K \cite{OGradyJMMM1996,CharapJAP1971,BaileyJAP1970}.  However, the permalloy oxide is a likely source of intermediate energy states in the barrier with random local magnetic environments that could easily contribute to loss of spin fidelity as initially spin-polarized electrons transport from Py to Al.  Importantly, our large $S_{s}/\alpha^{Py}_{\mathrm{abs}}$ values indicate that thermal injection suffers much less from this loss of signal due to interfacial effects.  

Though it is not possible to clearly identify a physical origin of this reduced sensitivity to the interface based on results presented here, we point out that the physical processes involved in electrical and thermal injection are potentially quite different.  This is particularly true when the clean interface limit is not achieved.  While electrical spin injection in this limit invokes tunneling of spin-polarized electrons, thermal injection in the tunneling limit could proceed by incoherent spin pumping as seen in the longitudinal spin Seebeck effect \cite{MeierNatCom2015,WuJAP2015,RezendePRB2014,UchidaPRX2014,UchidaJPCM2014,KehlbergerJAP2014,QuPRL2013,SchreierPRB2013,WeilerPRL2012}.  In this picture, the magnetic oxide could increase the effective interfacial spin mixing conductance or allow transport of spin via (non-electronic) collective spin excitations \cite{HahnEPL2014,WangPRL2014AF,ShiomiPRL2014,ZinkPRB2015,LucassenAPL2011}, and these effects could contribute to the SDSE signal measured here.  Further experiments exploring thermal spin injection in a range of materials and with more carefully controlled and characterized interfaces are required to clarify the potential advantages of thermal spin injection for a wide range of potential spintronic applications.

\section{Conclusions}

In summary, we presented evidence of thermally generated pure spin currents in permalloy/aluminum non-local spin valve structures.  Electrical spin injection, combined with contact resistance and using the actual geometry of the nanoscale devices determined from SEM images, indicated high resistance junctions and a reasonable fit of the spin signal to the 1d model of spin diffusion assuming tunneling contacts. The fit yielded very low values of spin polarization that we attribute to presence of oxidized permalloy that remains at the FM/NM interface.  Surprisingly, thermal spin injection remains efficient, suggesting that the oxidized permalloy participates in converting heat in the metallic FM into pure spin current in the NM, presumably via excitation of a collective magnetization.  We also briefly discussed challenges in quantifying thermal gradients in nanoscale structures, and described two methods for estimating thermal gradients in the NLSV.  We used these to quote a spin-dependent Seebeck coefficient in this Py/Al structure at $78$ K near $1\ \mu\mathrm{V}/\mathrm{K}$, which agrees well with previous reports on Py/Cu structures at $300$ K when compared as a fraction of the total absolute Seebeck coefficient.


\section{Acknowledgements}

We thank C. Leighton, P. Crowell, and L. O'Brien for many helpful discussions on fabrication and physics of NLSVs,  J. Aumentado for advice on EBL, G. C. Hilton, J. Beall, and J. Neibarger for advice and assistance on RF surface preparation and Al deposition, and J. Nogan and the IL staff at CINT for guidance and training.   We gratefully acknowledge support from the DU PROF program (A.H.) as well as the NSF (grants DMR-0847796 and DMR-1410247, D.W. and B.L.Z.).    B.L.Z.\ also thanks the University of Minnesota Chemical Engineering and Materials Science Department, as a portion of this work benefitted from support of the George T. Piercy Distinguished Visiting Professorship.  This work was performed, in part, at the Center for Integrated Nanotechnologies, an Office of Science User Facility operated for the U.S. Department of Energy (DOE) Office of Science by Los Alamos National Laboratory (Contract DE-AC52-06NA25396) and Sandia National Laboratories (Contract DE-AC04-94AL85000).




%

\appendix

\setcounter{equation}{0}
\setcounter{figure}{0}
\setcounter{table}{0}
\makeatletter
\renewcommand{\theequation}{A\arabic{equation}}
\renewcommand{\thefigure}{A\arabic{figure}}
\renewcommand{\thetable}{A\arabic{table}}
\renewcommand{\bibnumfmt}[1]{[S#1]}

\section{Appendix A: Analytic Thermal Modeling of NLSVs \label{AnaMod}}
For the case of electrical spin injection (Fig.\ \ref{NLSVcartoon}a) in steady state with $I$ applied to junction $1$, we can write two coupled equations for heat flow:
\begin{equation}
	\label{SteadyStateHotNLSV}
	 P_{J} +  P_{\Pi} = K_{\mathrm{Sub}}(T^{e}_{1} - T_0) + K_{\mathrm{nm}}(T^{e}_{1} - T^{e}_{2})
\end{equation}
\begin{equation}
	\label{SteadyStateColdNLSV}
	0 = K_{\mathrm{Sub}}(T^{e}_{2} - T_o) + K_{\mathrm{nm}}(T^{e}_{2} - T^{e}_{1}).
\end{equation}
Where $T^{e}_{1}$ ($T^{e}_{2}$) indicate the temperature of junction 1 (2) in response to power applied to junction 1 in the electrical spin injection configuration (Fig.\ \ref{NLSVcartoon}a). 
These can be solved to give the temperature differences between the junctions and the substrate:
\begin{eqnarray}
T^{e}_{2}-T_{0}&=& \frac{K_{\mathrm{nm}}(P_{J}+P_{\Pi}) }{K_{\mathrm{sub}}(K_{\mathrm{sub}}+2 K_{\mathrm{nm}}) },\\
               \Delta T^{e}_{2}      & = & \frac{K_{\mathrm{nm}}(I^{2}R^{e}_{\mathrm{eff}}+I\alpha_{\mathrm{rel},1}T_{0}) }{K_{\mathrm{sub}}(K_{\mathrm{sub}}+2 K_{\mathrm{nm}}) } \label{DeltaT2e}
\end{eqnarray}
and
\begin{eqnarray}
T^{e}_{1}-T_{0}& = & \frac{P_{J}+P_{\Pi}}{K_{\mathrm{Sub}}}-\Delta T^{e}_{2}\\
                  \Delta T^{e}_{1}   & = & \frac{I^{2}R^{e}_{\mathrm{eff}}+I\alpha_{\mathrm{rel},1}T_{0}}{K_{\mathrm{sub}}} -\Delta T^{e}_{2}.
\end{eqnarray}
This combination of Joule and Peltier power applied to junction 1 will lead to a voltage contribution from purely thermoelectric effects at junction 2, $V_{\mathrm{NLE}}=\alpha_{\mathrm{rel},2}\Delta T^{e}_{2}$.  Eq. \ref{DeltaT2e} clearly shows that this voltage will have terms $\propto$ both $I$ and $I^{2}$, as seen in Figs. \ref{PminAP}a) and \ref{IVs}b).   


Similar expressions describe the device in the thermal spin injection configuration (Fig.\ \ref{NLSVcartoon}c).  Here only Joule heating is expected, as shown in the thermal model schematic inset in Fig.\ \ref{IVs}c), so that when current is driven through FM1:
\begin{equation}
	\label{SteadyStateHotNLSVt}
	 P_{J}  = K_{\mathrm{Sub}}(T^{t}_{1} - T_o) + K_{\mathrm{nm}}(T^{t}_{1} - T^{t}_{2})
\end{equation}
\begin{equation}
	\label{SteadyStateColdNLSVt}
	0 = K_{\mathrm{Sub}}(T^{t}_{2} - T_o) + K_{\mathrm{nm}}(T^{t}_{2} - T^{t}_{1}).
\end{equation}
Here $T^{t}_{1}$ ($T^{t}_{2}$) indicate the temperature of junction 1 (2) in response to power applied to FM1 in the thermal spin injection orientation (Fig.\ \ref{NLSVcartoon}c). 

Again these can be solved to give the temperature differences between the junctions and the substrate:
\begin{eqnarray}
T^{t}_{2}-T_{0}&=& \frac{K_{\mathrm{nm}}(P_{J}) }{K_{\mathrm{sub}}(K_{\mathrm{sub}}+2 K_{\mathrm{nm}}) },\\
               \Delta T^{t}_{2}      & = & \frac{K_{\mathrm{nm}}(I^{2}R^{t}_{\mathrm{eff}}) }{K_{\mathrm{sub}} \label{DeltaTt2}(K_{\mathrm{sub}}+2 K_{\mathrm{nm}}) }
\end{eqnarray}
and
\begin{eqnarray}
T^{t}_{1}-T_{0}& = & \frac{P_{J}}{K_{\mathrm{Sub}}}-\Delta T^{t}_{2} \\
                  \Delta T^{t}_{1}   & = & \frac{I^{2}R^{t}_{\mathrm{eff}}}{K_{\mathrm{sub}}} -\Delta T^{t}_{2}.
\end{eqnarray}
The Joule power applied to FM1 will  again lead to a voltage contribution from purely thermoelectric effects at junction 2, $V_{\mathrm{NLT}}=\alpha_{\mathrm{rel},2}\Delta T^{t}_{2}$.  As expected, the model predicts only $\propto I^{2}$ terms for $V_{\mathrm{NLT}}$, and the measurements (Figs. \ref{PminAP}b) and \ref{IVs}c)) are indeed nearly perfect parabolas.   


Finally, we note that the ``contact resistance" measurement, where the voltage is measured at the FM strip used for current injection as shown in Fig. \ref{NLSVcartoon}d) will give the sum of potentially three voltages: a voltage drop caused by current flow across the actual interface between NM and FM1 (the traditional understanding of a contact resistance), a potential difference due to geometrical current spreading in the nanoscale circuit\cite{ObrienNatComm2014}, and a voltage from thermoelectric effects due to the temperature gradients produced in the structure.  This sum is then:
\begin{equation}
V_{C}=IR_{\mathrm{C}}+ V_{\mathrm{spread}}+ \alpha_{\mathrm{rel},1}\Delta T^{e}_{1}.
\end{equation}
The thermoelectric voltage includes both $I$ and $I^{2}$ terms, and as seen in Fig.\ \ref{IVs}d) these IV curves show clear non-linearity.  It will also be important to consider the size of the thermoelectric term $\propto I$ relative to the average apparent resistance in using these effective 3-terminal measurements to judge which form of the 1d spin diffusion equation to choose for analysis of the spin transport in the NLSV \cite{TakahashiPRB2003}.  In the NLSV devices shown here, the thermoelectric $\propto I$ term is small compared to the total signal (on order of $100$ nV for the measurement shown in Fig.\ \ref{IVs}d).


This model therefore provides expressions for three voltage measurements as a function of applied current with terms proportional to $I$ and to $I^2$ as shown in Eqs. \ref{FitEqVnle}-\ref{FitEqVc}, where the $A_{i}$, $B_{i}$, and $C_{i}$ coefficients result from fits to the measured $V$ as a function of $I$ as shown in Fig.\ \ref{IVs}..  Measurements and fitting of these three voltages allows determination of the temperature profile in the device.  


%
%
%

\section{Appendix B: Interface Quality, Spin Transport Models, and Signal Size\label{Interface}}

\begin{figure}
\includegraphics[width=3.38in]{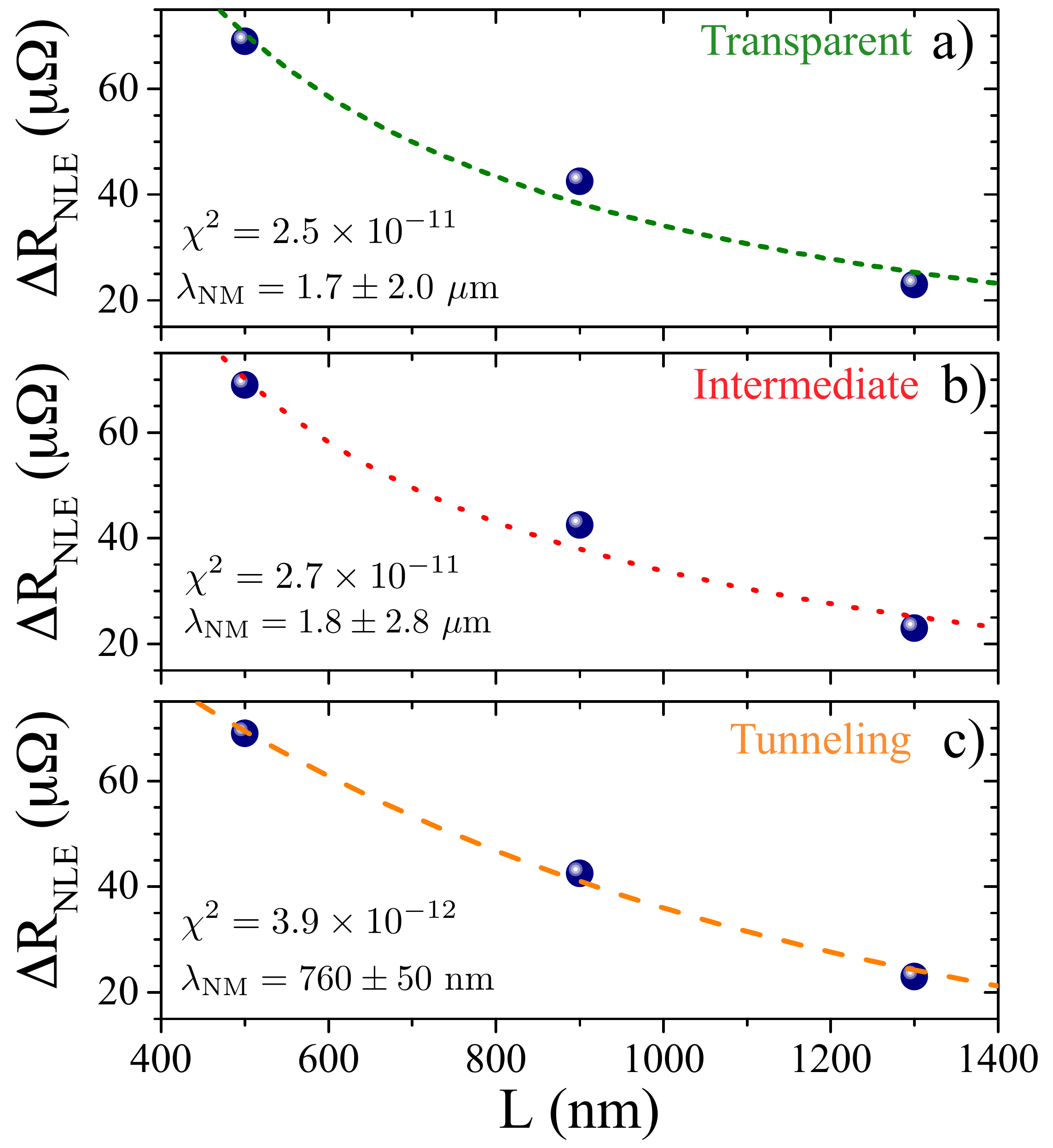}
\caption{Comparison of fits to the three limits of the 1d spin diffusion equation.  Fit to the tunneling model gives the lowest $\chi^{2}$ and best estimate of $\lambda_{\mathrm{NM}}$.}
\label{Limits}
\end{figure}

\begin{table}
\begin{tabular}{| | c | c | c | c | c ||}
\hline\hline
  Device & $L$ & $w_{\mathrm{FM_1}}\ \mathrm{(nm)} $& $w_{\mathrm{FM_2}}  \mathrm{(nm)}$ & $w_{\mathrm{NM}}\mathrm{(nm)}$  \\
 \hline
 $500\ \mathrm{nm}$  &  $475 $  & $190 $  & $400 $  & $510 $ \\
 $900\ \mathrm{nm}$  &  $850 $  & $230 $  & $415 $  & $485 $ \\
 $1300\ \mathrm{nm}$  &  $1260 $  & $225 $  & $425 $  & $460 $ \\
 \hline \hline

\end{tabular}
\caption{NLSV geometries as measured by scanning electron micrography.  Each dimension has an estimated error of $30$ nm. }
\label{Geometry}
\end{table}

The main result of our study is that thermal spin injection is more tolerant of, and perhaps enhanced by, an imperfect FM/NM injector interface.  In the main text we highlight this by comparison to Slachter et al.'s result based on an NLSV with a very different geometry, specifically a smaller area of contact between FM and NM.  If we had reached the transparent limit for the device geometry studied here, a significant reduction in the spin signal should be expected from spin backflow into the injecting FM.   However, we see no reason that such spin absorption would affect electrical spin injection differently than thermal spin injection, so that comparing the ratios is still appropriate.  

Nevertheless, close examination of which interface limit is most appropriate to interpret our results remains important.
As stated in the main text, our best determination of sample geometry suggests that $R_{c}>\mathcal{R}_{\mathrm{NM}}>\mathcal{R}_{\mathrm{FM}}$.  Takahashi and Maekawa \cite{TakahashiPRB2003}, originally derived the 1d spin transport equation in the tunneling limit for   $R_{c}\gg\mathcal{R}_{\mathrm{NM}}$, but clarify that the tunneling equation gives close values of $\Delta R_{s}$ even when $R_{c}\approx\mathcal{R}_{\mathrm{NM}}$ so that whenever $R_{c}\gtrsim \mathcal{R}_{\mathrm{NM}}$ the tunneling model should be used.  In Fig.\ \ref{Limits} we compare fits to the three different 1d spin models for our $\Delta R_{\mathrm{NLE}}$ data.  In addition to the tunneling eq. (Eq. \ref{1dSpin}), these are the transparent case ($R_{c}\ll \mathcal{R}_{\mathrm{FM}} $):
\begin{equation}
\Delta R_{\mathrm{s}}=4\frac{\alpha^{2}\mathcal{R}^{2}_{\mathrm{FM}}}{(1-\alpha^{2})\mathcal{R}_{\mathrm{NM}}}\frac{e^{\left( -L/\lambda_{\mathrm{NM}}\right)}}{ \left[1+ \frac{2 \mathcal{R}_{\mathrm{FM}}}{(1-\alpha^{2}) \mathcal{R}_{\mathrm{NM}}} \right]^{2} - e^{ \left( -2L/\lambda_{\mathrm{NM}}\right)}},
\label{transparent}
\end{equation}
and the intermediate case ($\mathcal{R}_{\mathrm{NM}} \gg R_{c} \gg \mathcal{R}_{\mathrm{FM}}$):
\begin{equation}
\Delta R_{\mathrm{s}}=4 \frac{P^{2}_{I}}{\left( 1- P^{2}_{I}\right)^{2}} \frac{R_{c}^{2}}{\mathcal{R}_{\mathrm{NM}}} \frac{e^{-L/\lambda_{\mathrm{NM}}}}{1-e^{-2L/\lambda_{\mathrm{NM}}}}.
\label{intermediate}
\end{equation}

Note that calculation of $\Delta R_{s}$ in the transparent model (for our NLSV geometry assuming typical values of $\alpha=0.38$) give $\Delta R_{s}\sim 150-300\ \mathrm{\mu \Omega\ cm}$ depending on if the additional sidewall FM/NM interface is taken into account.    This is still much larger than our measured electrical spin signals.
 
In each panel, the corresponding fit is shown along with the resulting goodness-of-fit measure, $\chi^{2}$, and the fitted $\lambda_{\mathrm{NM}}$ with associated error estimate.  For this device, it is clear that the tunneling model gives the best $\chi^{2}$ and also is the only fit that leads to physically sensible values of $\lambda_{\mathrm{NM}}$.  Fixing $\lambda_{\mathrm{NM}}$ to the lower values from the tunneling fit that are more in line with literature led to poorer fits.  We also note that this fit was performed with the nominal $L$ values, but use of $L$ measured from SEM imaging of the NLSV devices after testing (which are close to the nominal values for the case of $L$) give similar fits with parameters within error bars of those shown here. This fitting, combined with the evidence from the magnitude of contact and spin resistances, suggests the tunneling model is the most appropriate choice for these NLSV devices, and leads us to conclude that the strongly reduced $P_{I}$ most likely results from intermediate states in the barrier.   

Finally, we point out that particularly since the NM channel in these NLSV devices is wider than intended, use of a 3d spin transport model could offer benefits for understanding the balance between electrical and thermal spin injection in such a geometry \cite{KimuraJPCM2007,HamrlePRB2005}. In future studies we intend to produce narrower channels via improved lithography, but will also explore implementation of the more complicated 3d models.

\end{document}